\documentclass[12pt]{iopart}

\usepackage{graphicx}%

\begin{document}

\title{Evidence of magnetic mechanism for cuprate superconductivity}

\author{Amit Keren}

\address{Physics Department, Technion-Israel Institute of Technology, Haifa
32000, Israel}

\ead{keren@physics.technion.ac.il}

\begin{abstract}
A proper understanding of the mechanism for cuprate superconductivity can
emerge only by comparing materials in which physical parameters vary one at
a time. Here we present a variety of bulk, resonance, and scattering
measurements on the (Ca$_{x}$La$_{1-x}$)(Ba$_{1.75-x}$La$_{0.25+x}$)Cu$_{3}$O%
$_{y}$ high temperature superconductors, in which this can be done. We
determine the superconducting, N\'{e}el, glass, and pseudopage critical
temperatures. In addition, we clarify which physical parameter varies, and,
equally important, which does not, with each chemical modification. This
allows us to demonstrate that a single energy scale, set by the
superexchange interaction $J$, controls all the critical temperatures of the
system. $J$, in-turn, is determined by the in plane Cu-O-Cu buckling angle.
\end{abstract}


\maketitle

\section{Introduction}

\begin{figure}
[ptb]
\begin{center}
\includegraphics[
natheight=6.677200in,
natwidth=8.531400in,
height=2.5538in,
width=3.2569in
]%
{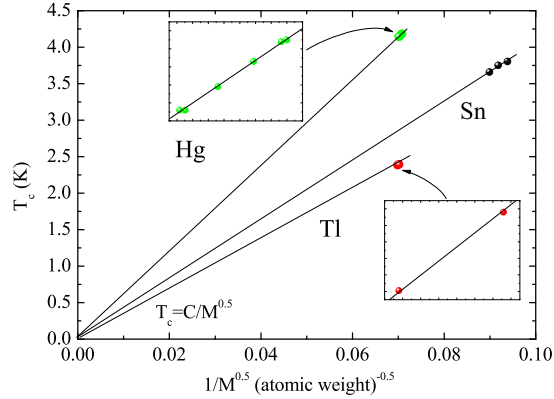}%
\caption{The superconducting critical temperature $T_{c}$ of three metallic
superconductors as a function if the inverse root of their mass \cite{Maxwell}%
. The solid lines are strait. The insets are zoom on the data.}%
\label{isotope}%
\end{center}
\end{figure}

The critical temperature for superconductivity $T_{c}$ in the metallic
superconductors Hg, Sn, and Tl as a function of $M^{-1/2}$, where $M$ is the
atomic mass, is presented in Fig.~\ref{isotope} on a full scale including
the origin \cite{Maxwell}. In the case of Sn a clear isotope effect is
observed resulting in a 4\% variation of $T_{c}$ upon isotope substitution.
For Hg and Tl the variations are observed only by zooming in on the data. In
all cases a linear fit goes through the data points and the origin quite
satisfactorily, namely, $T_{c}$ is proportional to $M^{-1/2}$. This
observation, known as the isotope effect, plays a key role in exposing the
mechanism for superconductivity in metallic superconductors. However, had
nature not provided us with isotopes, and we had to draw conclusions only by
comparing the different materials in Fig.~\ref{isotope}, we would conclude
that $T_{c}$ has nothing to do with the atomic mass. Thus, Fig.~\ref{isotope}
demonstrates that it is dangerous to compare materials where several
quantities vary simultaneously. The isotope experiment overcomes this
problem and reveals the origin of metallic superconductivity.

The mechanism for high temperature superconductivity (HTSC) in the cuprate
is still elusive, but is believed to be of magnetic origin \cite%
{Belivers,KotliarPRB88}. Verifying this belief would require an experiment
similar to the isotope effect, namely, a measurement of $T_{c}$ versus the
magnetic interaction strength $J$, with no other structural changes in the
compounds under investigation. Unfortunately, varying $J$ experimentally can
only be done by chemical variation, usually leading to very different
materials. For example, YBa$_{2}$Cu$_{3}$O$_{y}$ (YBCO) has a maximum $T_{c}$
of 96~K, La$_{2-x}$Sr$_{x}$CuO$_{4}$ has a maximum $T_{c}$ of 38~K, and both
have roughly the same $J$ \cite{TranquadaPRB40,KeimerPRB92,Wan08105216}.
This fact has been used to contradict the magnetic mechanism although these
materials are different in crystal perfection, number of layers, symmetry,
and more. Clearly they are uncomparable exactly as Hg, Sn and Tl.

In the present manuscript we describe a set of experiments designed to
overcome this problem and to perform a magnetic analog of the isotope
experiment by making very small and subtle chemical changes, which modify $J$
but keep all other parameters intact. This is achieved by investigating a
system of HTSC with the chemical formula (Ca$_{x}$La$_{1-x}$)(Ba$_{1.75-x}$La%
$_{0.25+x}$)Cu$_{3}$O$_{y}$ and acronym CLBLCO. Each value of $%
x=0.1\ldots0.4 $ is a family of superconductors. All families have the YBCO
structure with negligible structural differences; all compounds are
tetragonal, and there is no oxygen chain ordering as in YBCO. Within a
family, $y$ can be varied from zero doping to over doping.

We present measurements of the critical temperature of superconductivity $%
T_{c}$ using resistivity \cite{Goldschmidt}, the spin glass temperature $%
T_{g}$ \cite{KanigelPRL02} and the N\'{e}el temperature $T_{N}$ \cite%
{OferPRB06} of the parent antiferromagnet (AFM) using zero field muon spin
relaxation ($\mu$SR), the level of doping and the level of impurities by
Nuclear Quadruple Resonance (NQR) \cite{KerenPRB06}, the superconducting
carrier density using transverse field muon spin rotation \cite{KerenSSC03},
the lattice parameters, including the oxygen buckling angle, with neutron
scattering \cite{OferPRB08}, and the pseudogap (PG) temperature $T^{\ast}$
with susceptibility \cite{LubaPRB08}. This allowed us to generate one of the
most complete phase diagrams of any HTSC system, to demonstrate a
proportionality between $T_{c}$ and $J$, and to draw more conclusions.

The paper consists of two main sections in addition to this introduction: in
Sec.~\ref{Main} the main experimental results and parameters extracted from
the raw data are presented. In Sec.~\ref{Conc} the conclusions are
summarized. Experimental details, raw data, and analysis description are
given in appendices.

\section{Main Results\label{Main}}

\begin{figure}
[ptb]
\begin{center}
\includegraphics[
natheight=6.536300in,
natwidth=8.445800in,
height=2.316in,
width=2.9871in
]%
{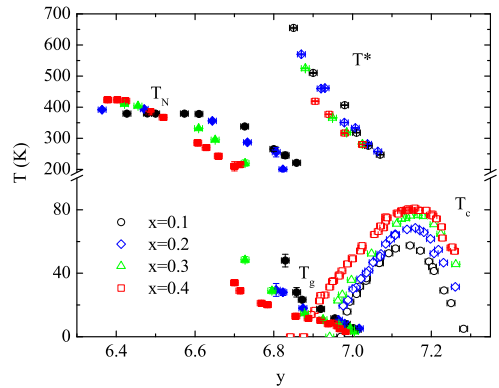}%
\caption{The phase diagram of the (Ca$%
_{x} $La$_{1-x}$)(Ba$_{1.75-x}$La$_{0.25+x}$)Cu$_{3}$O$_{y}$ system showing
the N\'{e}el ($T_{N}$), glass ($T_{g}$), superconducting ($T_{c}$), and
pseudogap ($T^{\ast}$) temperatures over the full doping range.}%
\label{criticalvsy}%
\end{center}
\end{figure}

The phase diagram of CLBLCO including $T_{N}$, $T_{g}$, $T_{c}$, and $%
T^{\ast }$ versus oxygen level $y$ is shown in Fig.~\ref{criticalvsy}.
Details of the magnetic measurements are given in ~\ref{MagCritical}
and pseudogap measurements in ~\ref{PG}. In the doping region up to $%
y=6.5$, the $T_{N}$ curves of the different families are nearly parallel.
The maximum $T_{N}$ ($T_{N}^{max}$) of the $x=0.1$ family is the lowest, and
of the $x=0.4$ family the highest. Upon further doping $T_{N}$ decreases
rapidly and differently, leading to a crossing point after which the $x=0.4$
family has the lowest value of $T_{N}$, and the $x=0.1$ family the highest.
By further doping, the long-range order is replaced by a spin glass phase,
where islands of spins freeze. The spin glass phase penetrates into the
superconducting phase which exists for $y=6.9$ to $y=7.25$. This phase
starts earlier as $x$ increases. The superconducting domes are nearly
concentric with maximum $T_{c}$ ($T_{c}^{max}$) decreasing with decreasing $%
x $. $T_{c}^{max}$ varies from 80 K at $x=0.4$ to 56 K at $x=0.1$, a nearly
30\% variation. This is much stronger varation than the strongest isotope
effect in nature. As for the pseudogap temperatures $T^{\ast}$, it seems
that the $x=0.1$ family has the highest $T^{\ast}$ and the $x=0.4$ the
lowest.

Perhaps the clearest feature of this phase diagram is the correlation
between $T_{N}^{max}$ and $T_{c}^{max}$. The family with the highest $%
T_{N}^{max}$ has the highest $T_{c}^{max}$. However, $T_{N}$ is not a clean
energy scale. It is well established that a pure 2D AFM orders magnetically
only at $T=0$, and that $T_{N}$ is finite only for 3D AFM. Intermediate
cases are described by more complicated interactions where $J$ is the
isotropic intralayer Heisenberg interaction, and $\alpha_{eff}J$ represent interlayer
and anisotropic coupling \cite{KeimerPRB92}. In order to extract $J$ from $%
T_{N}$, $\alpha_{eff}$ must be determined.

\begin{figure}
[ptb]
\begin{center}
\includegraphics[
natheight=5.827100in,
natwidth=7.443400in,
height=2.3791in,
width=3.0338in
]%
{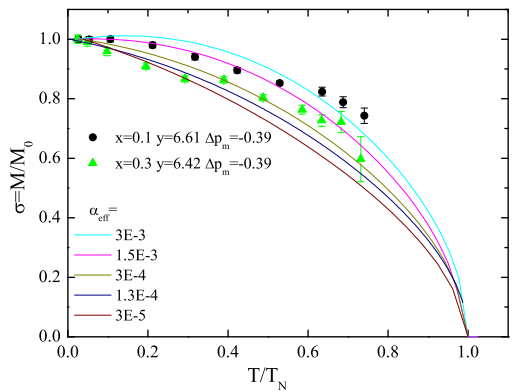}%
\caption{The normalized staggered
magnetization as a function of the normalized temperature. The symbols are
the experimental results for two CLBLCO samples, taken by measuring the
oscillation frequency of the muon polarization curves. The parameter $\Delta
p_{m}$ is explained in the text. The solid lines are the theoretical curves
for various effective anisotropic and interlayer coupling $\protect\alpha%
_{eff}J$ \protect\cite{OferPRB06}.}%
\label{alpha}%
\end{center}
\end{figure}

One method of extracting $\alpha_{eff}$ is from the magnetic order parameter 
$M$ versus temperature $T$ \cite{KeimerPRB92,ArovasPRB98}. For small $%
\alpha_{eff}$ the reduction of the magnetic order parameter $M$ with
increasing $T$ is fast so that at $\alpha_{eff}=0$ the 2D limit is
recovered. On the other hand, in the three dimensional case, where $%
\alpha_{eff}=1$, we expect a weak temperature dependence of $M$ at $%
T\rightarrow0$ due to lack of antiferromagnetic magnons states at low
frequencies. A plot of the normalized order parameter $\sigma=M/M_{0}$,
where $M_{0}$ is the order parameter at $T\rightarrow0$, versus $T/T_{N}$
should connect (1,0) to (0,1) as depicted in Fig.~\ref{alpha} \cite%
{OferPRB06}. The differences between curves are set only by $\alpha_{eff}$
and they can be fitted to experimental data. Given $\alpha_{eff}$ and $T_{N}$%
, $J$ can be extracted. This is not a very accurate method of $\alpha_{eff}$
determination, but $J$ depends only on $\ln(\alpha_{eff})$ so high accuracy
is not required \cite{KeimerPRB92}.

Using zero field muon spin rotation, we determined the muon angular rotation
frequencies $\omega$ in the different compounds \cite{OferPRB06}. The
normalized order parameter is given by $\sigma(T)=\omega(T)/\omega(0)$. The
order parameter extracted from the high angular frequency, around a few tens
of MRad/Sec ($\omega\sim27$ MRad/Sec in our case), is known to agree with
neutron scattering determination of $\sigma$.

In Fig.~\ref{alpha} we also present $\sigma$ for two different underdoped
CLBLCO samples with $x=0.1$ and $0.3$. Clearly the reduction of the
magnetization with increasing temperatures is not the same for these two
samples, and therefore their anisotropies are different. Since $\sigma$ is
less sensitive to increasing $T$ in the $x=0.1$ family than in the $x=0.3$
family, the $\alpha_{eff}$ of $x=0.1$ must be larger. Using the muon spin
rotation frequency versus $T$ and calculations based on the Swinger boson
mean field (SBMF) theory \cite{OferPRB06,KeimerPRB92}, we determined $%
\alpha_{eff}$ for all samples \cite{OferPRB06}. Knowing $T_{N}$ and $%
\alpha_{eff}$ for all samples we extract a corrected $T_{N}$ ($T_{N}^{cor}$%
). For the very underdoped samples this $T_{N}^{cor}=J$. For doped samples
the situation is more complicated since the samples are described more
accurately by the t-J model. We will present $T_{N}^{cor}$ shortly after
discussing the doping.

\begin{figure}
[ptb]
\begin{center}
\includegraphics[
natheight=8.461300in,
natwidth=6.438500in,
height=2.9888in,
width=2.2805in
]%
{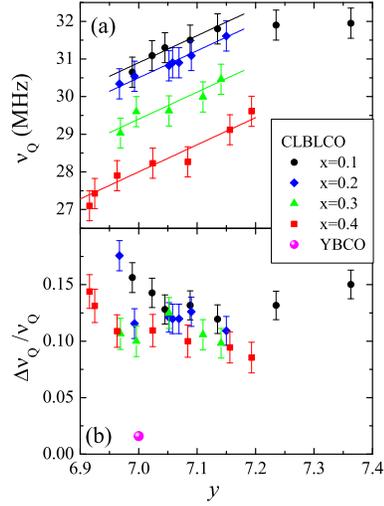}%
\caption{(a) Nuclear quadrupole
resonance frequncy and (b) line width of the in plane $^{63}$Cu(2) in all
the CLBLCO samples extracted from NMR spetra \protect\cite{KerenPRB06}. Data
for YBa$_{2}$Cu$_{3}$O$_{7}$ are also shown.}%
\label{nqranalysis}%
\end{center}
\end{figure}

Doping in CLBLCO is done by controlling the oxygen level in a chain layer as
in YBCO. This leaves some ambiguity concerning the doping of the CuO$_{2}$
planes. One possibility to determine the amount of charge present in this
plane is to measure the in-plane Cu NQR frequency $\nu_{Q}$. Assuming the
lattice parameters variations within a family can be ignored (an assumption
tested with neutrons in ~\ref{Lattice}) the NQR frequency is
proportional to the level of doping in the plane. The in-plane Cu NQR
frequencies, discussed further in ~\ref{Impurities}, are shown in
Fig.~\ref{nqranalysis}(a). It is clear that $\nu_{Q}$ grows linearly with
doping in the underdoped side of the phase diagram. The most interesting
finding is that, within the experimental error, the slope of $\nu_{Q}(x,y)$
in the underdoped side is $x$-independent, as demonstrated by the parallel
solid lines. This means that the rate at which holes $p$ are introduced into
the CuO$_{2}$ planes, $\partial p/\partial y$, is a constant independent of $%
x$ or $y$ in the underdoped region. Using further the ubiquitous assumption
that the optimal hole density, at optimal oxygenation, $y_{opt}$, is
universal, we conclude that the in-plane hole density is a function only of $%
\Delta y=y-y_{opt}$. The same conclusion was reached by X-ray absorption
spectroscopy (XAS) experiments \cite{SannaXray}.

In contrast, the CLBLCO family obeys the Uemura relations in the entire
doping region, namely, $T_{c}$ is proportional to $n_{s}/m^{\ast }$ where $%
n_{s}$ is the superconducting carrier density and $m^{\ast }$ the effective
mass \cite{KerenSSC03}. This is determined with transverse field muon spin
relaxation measurements where the Gaussian relaxation rate $R_{\mu }$ is
proportional to $n_{s}/m^{\ast }$ as explained in ~\ref{PenDep} \cite%
{MuonBook,SonierReview}. The experimental results are depicted in Fig.~\ref%
{Uemura} for all families. This experiment seems to contradict the NQR
results for the following reason. If all holes had turned superconducting ($%
p=n_{s}$), then samples of different $x$ but identical $\Delta y$ should
have identical $\Delta p$ and identical $\Delta n_{s}$. In addition, if $%
m^{\ast }$ is universal, samples with a common $\Delta y$ should have the
same $T_{c}$, in contrast to the phase diagram of Fig.~\ref{criticalvsy}.
Something must be wrong in the hole counting. A similar conclusion was
reached in the investigation of Y$_{1-x}$Ca$_{x}$Ba$_{2}$Cu$_{3}$O$_{6+y}$ 
\cite{SannaCM}.

\begin{figure}
[ptb]
\begin{center}
\includegraphics[
natheight=6.327800in,
natwidth=7.944200in,
height=2.4794in,
width=3.1055in
]%
{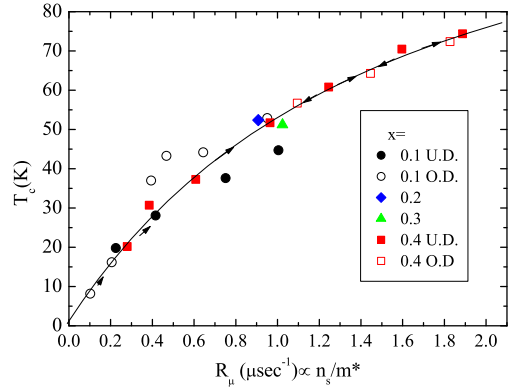}%
\caption{A Uemura plot showing $T_{c}$
versus the muon relaxation rate $R_{\protect\mu}$ expected to be
proportional to the superconducting carrier density $n_{s}$ over the
effective mass $m^{\ast}$ for the CLBLCO family of superconductors.}%
\label{Uemura}%
\end{center}
\end{figure}

\begin{figure}
[ptb]
\begin{center}
\includegraphics[
natheight=7.988300in,
natwidth=6.411700in,
height=2.9897in,
width=2.4042in
]%
{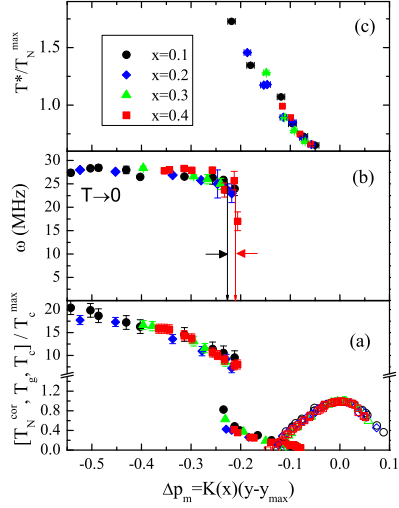}%
\caption{Unified phase diagram of CLBLCO
including several experimental parameters ploted as a function of mobile
hole variation $\Delta p_{m}$ as described in the text: (a) superconducting
critical temperature $T_{c}$, the spin glass temperature $T_{g}$, and the
corrected N\'{e}el temperature $T_{N}^{cor}$ which is identical to the
Heisenberg supperexchange $J$ at zero doping. All parameters are normalized
by $T_{c}^{\max }$ of each family.(b) The zero temperature muon oscillation
frequency equivalent to the magnetic order parameter, for all four CLBLCO
families. The vertical lines show the expected difference in the critical
doping had it vary by 5\% between families. (c) The pseudogap temperature
normalized by the maximum Neel temperature (see text) $T^{\ast }/T_{N}^{max}$%
.}%
\label{unified}%
\end{center}
\end{figure}

This problem can be solved by assuming that not all holes are mobile and
turn superconducting. Therefore, we should define the mobile hole
concentration $\Delta p_{m}$ by multiplying $\Delta y$ by a different
constant per family $K(x)$, namely, $\Delta p_{m}=K(x)\Delta y$. The
superconducting carrier density variation $\Delta n_{s}$ is now proportional
to $\Delta p_{m}$ with a universal factor. The $K$s are chosen so that the
superconducting critical temperature $T_{c}$ domes, normalized by $%
T_{c}^{max}$ of each family, collapse onto each other. This is shown in Fig.~%
\ref{unified}(a) using $K=0.76$, $0.67$, $0.54$ ,$0.47$ for $x=0.1\ldots0.4$%
. An animation showing the rescaling of the critical temperatures and doping
are given in the supporting materials.

Figure \ref{unified}(a) also shows the other critical temperatures $%
T_{N}^{cor}$ and $T_{g}$ normalized by $T_{c}^{max}$ and plotted as a
function of $\Delta p_{m}$. The critical temperatures from all families
collapse to a single function of $\Delta p_{m}$. This means that there is
proportionality between the in-plane Heisenberg coupling constant $J$ and
the maximum of the superconducting transition temperature $T_{c}^{max}$ in
the series of (Ca$_{x}$La$_{1-x}$)(Ba$_{1.75-x}$La$_{0.25+x}$)Cu$_{3}$O$_{y}$
families. In fact, the data presented up to here can be explained by
replacing the $1/m\ast$ in the Uemura relation by a family-dependent
magnetic energy scale $J_{x}$ and writing 
\begin{equation}
T_{c}=cJ_{x}n_{s}(\Delta y)  \label{TcvsJandns}
\end{equation}
where $c$ is a universal constant for all families. For a typical
superconductor having $T_c=80$~K, $J\sim1000$~K, and $8~$\% superconducting
carrier density per Cu site, $c$ is on the order of unity. This is the main
finding of this paper. Theoretical indications for the importance of $J$ and 
$n_{s}$ in setting up $T_{c}$ could be found since the early days of HTSC 
\cite{KotliarPRB88}.

Having established a proportionality between $T_{c}$ and $J$, it is
important to understand the origin of the $J$ variations between families in
CLBLCO. As we show in appendix~\ref{Lattice} using lattice parameters
measurements with neutron diffraction, the Cu-O-Cu buckling angle is
responsible for these $J$ variations since it is the only lattice parameter
that shows strong differences between the families; there is an about 30\%
change from the $x=0.1$ family to $x=0.4$ \cite{OferPRB08}. This change is
expected since as $x$ increases, a positive charge is moving from the Y to
the Ba site of the YBCO structure, pulling the oxygen toward the plane and
flattening the Cu-O-Cu bond.

\begin{figure}
[ptb]
\begin{center}
\includegraphics[
natheight=7.959700in,
natwidth=6.454100in,
height=2.9888in,
width=2.4267in
]%
{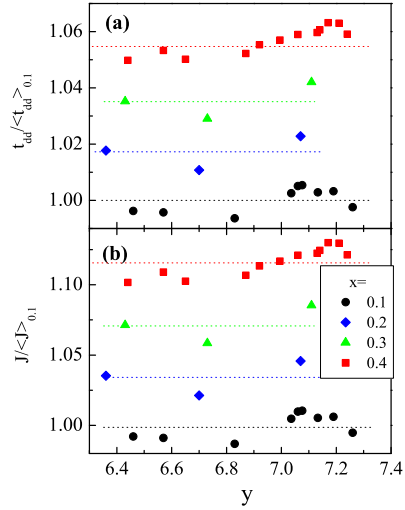}%
\caption{The hopping rate $t_{dd}$ (a),
and the superexchange coupling $J$ (b) obtained from crystal structure data
and the approximations $t_{dd}\propto cos\protect\theta/a^{-7}$ and $%
J\propto $ $t_{dd}^{2}$ where $\protect\theta$ is the buckling angle and $a$
the bond length shown in Fig.~\protect\ref{neutrons}. }%
\label{tjparam}%
\end{center}
\end{figure}

From the lattice parameters it is possible to construct the hopping integral 
$t$ and super-exchange $J$ of the t-J model, assuming that the Hubbard $U$ and the charge
transfer energy $\Delta$ are family-independent. The basic quantity is the
hopping integral t$_{pd}$ between a Cu $3d_{x%
{{}^2}%
-y%
{{}^2}%
}$ and O $2p$ orbitals \cite{ZaaneannJPhys87}. This hopping integral is
proportional to bond length $a$ to the power -3.5 \cite{HarrisonBook}. The
hopping from the O $2p$ to the next Cu $3d_{x%
{{}^2}%
-y%
{{}^2}%
}$ involves again the bond length and cosine of the buckling angle $\theta$.
Thus, the Cu to Cu hopping depends on $a$ and $\theta$ as $t_{dd}\propto
cos\theta/a^{-7}$ and $J$ is proportional to $t_{dd}%
{{}^2}%
$, hence, $J\propto cos%
{{}^2}%
\theta/a^{14}$.

Estimates of the $t_{dd}$ and $J$, normalized to the averaged values of the $%
x=0.1$ family, $\left\langle t_{dd}\right\rangle _{0.1}$ and $\left\langle
J\right\rangle _{0.1}$ are presented in Fig.~\ref{tjparam}(a) and (b).
Although there is a variation in $t$ and $J$ within each family, the
variation is much larger between the families. $J$ increases with increasing 
$x$, in qualitative agreement with experimental determination of $J$. There
is a 10\% increase in $J$ which is \emph{not} big enough to explain the
variations in $T_{N}^{cor}$, but the magnitude and direction are
satisfactory. More accurate calculations are on their way \cite{LePetit}. It
is also important to notice that there is an about 5\% difference in the $%
t/J $ ratio between the two extreme families.

Equally important is to understand what is not changing between families.
For example, we would like to check whether the crystal quality is the same
for all families. Figure \ref{nqranalysis}(b) shows the width of the NQR
lines discussed in ~\ref{Impurities} \cite{KerenPRB06}. This width
is minimal at optimal doping and is identical within experimental resolution
for all families. More recent experiments with oxygen 17 lines, and better
resolution, provide the same conclusions \cite{AmitInPrep}. A third evidence
for the similarity of crystal quality comes from XAS \cite{SannaPC}. Thus,
the variations in $T_{c}$ for the different families cannot be explained by
disorder or impurities.

The unified phase diagram in Fig.~\ref{unified}(a) reveals more information
about the CLBLCO family than just Eq.~\ref{TcvsJandns}. In particular, the
fact that the N\'{e}el order is destroyed for all families at the same
critical $\Delta p_{m}$ is very significant. The disappearance of the
long-range N\'{e}el order and its replacement with a glassy ground state
could be studied more clearly by following the order parameter as a function
of doping. Naturally, $\omega(T\rightarrow0)$ disappears (drops to zero)
only when the N\'{e}el order is replaced by the spin glass phase as seen in
Fig.~\ref{unified}(b). We found that the order parameter is universal for
all families, and in particular the critical doping is family-independent 
\cite{OferPRB08}. To demonstrate this point we show, using the two arrows in
Fig.~\ref{unified}(b), what should have been the difference in the critical
doping had it changed between the $x=0.4$ and $x=0.1$ by 5\% of the doping
from zero, namely, 5\% of $\Delta p_{m}=0.3$. Our data indicates that $%
M_{0}(\Delta p_{m})$ is family-independent to better than 5\%. This is a
surprising result considering the fact that $t/J$ varies between families by
more than 5\% and that the critical doping is expected to depend on $t/J$ 
\cite{MvsP}. A possible explanation is that the destruction of the AFM order
parameter should be described by a hopping of boson pairs where $t$ is
absorbed into the creation of tightly bound bosons leaving a prominent
energy scale $J$ \cite{HavilioPRL98}. The proximity of the magnetic critical
doping to superconductivity makes this possibility appealing.

Finally we discuss the scaling of the pseudogap temperature. $T^{\ast}$
determined by magnetization measurements (See ~\ref{PG}) behaves
like the well-known PG or the spin gap measured by other techniques on a
variety of superconductors samples \cite{PGReviews}. More importantly, a
small but clear family dependence of $T^{\ast}$ is seen. At first glance in
Fig.~\ref{criticalvsy}, it appears that $T^{\ast}$ has anti-correlation with 
$T_{c}^{max}$ or the maximum $T_{N}$ ($T_{N}^{max}$). The $x=0.4$ family,
which has the highest $T_{c}^{max}$ and $T_{N}^{max}$, has the lowest $%
T^{\ast}$, and vice versa for the $x=0.1$ family.

However, this conclusion is reversed if instead of plotting the $T^{\ast}$
as a function of oxygen level, it is normalized by $T_{N}^{max}$, and
plotted as a function of mobile hole variation $\Delta p_{m}$ \cite%
{LubaPRB08}. This is demonstrated in Fig.~\ref{unified}(c). Here $%
T_{N}^{max} $ are chosen so that the $T_{N}(\Delta p_{m})/T_{N}^{max}$
curves collapse onto each other, and are 379, 391.5, 410, and 423~K for the $%
x=0.1\ldots0.4$ families, respectively. Therefore, $T_{N}^{max}$ should be
interpreted as the extrapolation of $T_{N}$ to the lowest $\Delta p_{m}$.
Normalizing $T^{\ast}$ by $T_{c}^{max}$ does not provide as good data
collapse as the normalization by $T_{N}^{max}$ \cite{LubaPRB08}. We conclude
that a PG does exist in CLBLCO and that it scales with the maximum N\'{e}el
temperature of each family. Therefore the PG is a 3D phenomenon involving
both in- and out of-plane coupling. A similar conclusion was reached by
resistivity analysis \cite{SuPRB06} and theoretical considerations \cite%
{MillisPRL93}.

\section{Conclusions\label{Conc}}

In this work four families of cuprate superconductors with a maximum $T_{c}$
variation of $30$\% are investigated. It is demonstrated experimentally that
these families are nearly identical in their crystal structure and crystal
quality. The only detectable property that varies considerably between them
is the Cu-O-Cu buckling angle. This angle is expected to impact the holes
hopping rate $t$, hence, the magnetic super-exchange $J$ between Cu spins. $%
J $ in turn sets the scale for the N\'{e}el temperature where long range
antiferromagnetic order is taking place. Independent measurements of $J$
show that indeed $J$ varies between families and that $T_{c}$ grows when $J$
increases. A linear transformation from oxygen concentration to mobile hole
concentration can generate a unified phase diagram in which $T_{c}$ is in
fact proportional to $J$ for all doping. Since $T_{c}$ is also proportional
to the superconducting carrier density it obeys Eq.~\ref{TcvsJandns}.

Surprisingly, the critical density, where the N\'{e}el order is destroyed at
zero temperature upon doping, is identical for all families. The critical
doping is expected to depend on $t/J$. This result has two implications. On
the one hand it supports the validity of the linear doping transformation;
on the other it suggests that this transformation has eliminated $t$ from
the low temperature effective Hamiltonian.

Finally, it is found that the pseudogap temperature $T^{\ast}$, as measured
by susceptibility, scales better with the N\'{e}el temperature than with $J$
(or $T_{c}$). This suggests that $T^{\ast}$ is determined by both in- and
out-of-plane coupling, and should be viewed as a temperature where the
system attempts unsuccessfully to order magnetically.

\section{Acknowledgements}

The author acknowledges very helpful discussions with A. Kanigel, R. Ofer,
E. Amit, A. Auerbach, Y. J. Uemura, and H. Alloul. Financial support from
the Israel Science Foundation is also acknowledged.

\appendix

\section{Magnetic critical temperatures\label{MagCritical}}

\begin{figure}
[ptb]
\begin{center}
\includegraphics[
natheight=8.363600in,
natwidth=6.340800in,
height=2.9897in,
width=2.2736in
]%
{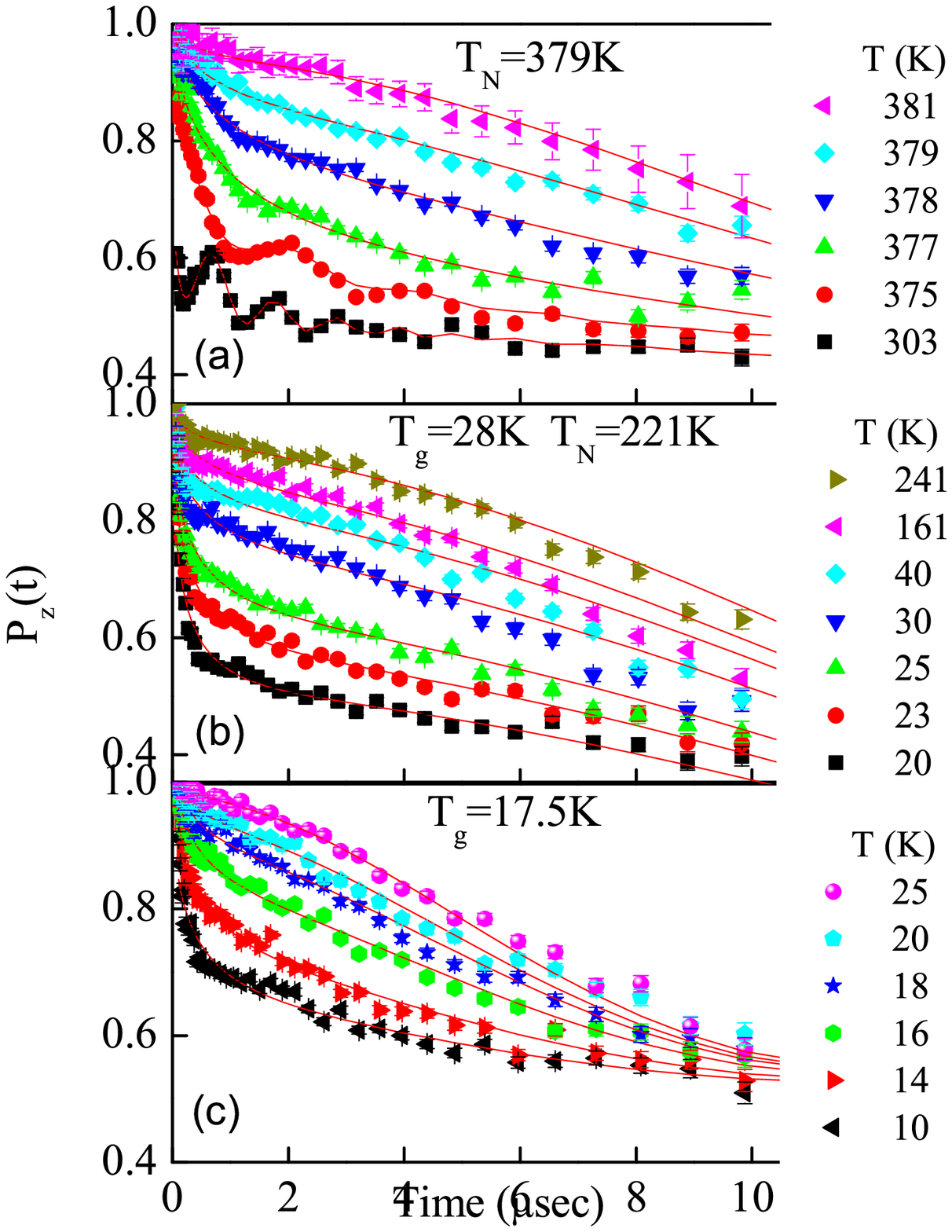}%
\caption{Time evolution of the muon
polarization for the $x=0.1$ family close to the magnetic critical
temperature. (a) A sample with an AFM transition. (b) A sample with both an
AFM and a spin-glass transition. (c) A sample with a spin-glass transition.
The solid lines are a fit as described in the text.}%
\label{musrzfraw}%
\end{center}
\end{figure}

The N\'{e}el and spin glass temperatures presented in Figs.~\ref{criticalvsy}%
, \ref{alpha} and \ref{unified} are obtained by zero field $\mu$SR. In these
experiments we determine the time-dependent spin polarization $P_{z}(t)$ of
a muon injected into the sample at different temperatures. $z$ represents
the initial muon spin direction. Figure \ref{musrzfraw} shows typical $%
P_{z}(t)$ curves, at different temperatures, for three samples from the $%
x=0.1$ family. At high temperatures the polarization curves from all samples
are typical of magnetic fields emanating from nuclear magnetic moments. In
this case the time-dependence of the polarization exhibits a Gaussian decay.
As the temperature is lowered the sample enters a magnetic frozen phase and
the polarization relaxes much more rapidly. While the transition from the
paramagnetic to the frozen state looks identical for all samples, the
behavior at very low $T$ is different and indicates the nature of the ground
state. Figure \ref{musrzfraw}(a) is an example of an antiferromagnetic
ground state. When the temperature decreases, long range magnetic order is
established at $\sim377$ K reflected by spontaneous oscillations of $P(t)$.

Figure \ref{musrzfraw}(c) is an example of a spin glass (SG) transition at $%
\sim17$ K. In this case the ground state consists of magnetic islands with
randomly frozen electronic moments \cite{KanigelPRL02}, and consequently the
polarization shows only rapid relaxation. When the transition is to a N\'{e}%
el or spin glass state, the critical temperatures are named T$_{N}$ and T$%
_{g}$, respectively. Figure \ref{musrzfraw}(b) presents an intermediate case
where the sample appears to have two transitions. The first one starts below
240 K, where the fast decay in the polarization appears. Between 160 K and
40 K there is hardly any change in the polarization decay, and at 30 K there
is another transition manifested in a faster decaying polarization. This
behavior was observed in all the samples on the border between
antiferromagnet and spin-glass in the phase diagram.

\begin{figure}
[ptb]
\begin{center}
\includegraphics[
natheight=5.771800in,
natwidth=8.015100in,
height=1.932in,
width=2.6766in
]%
{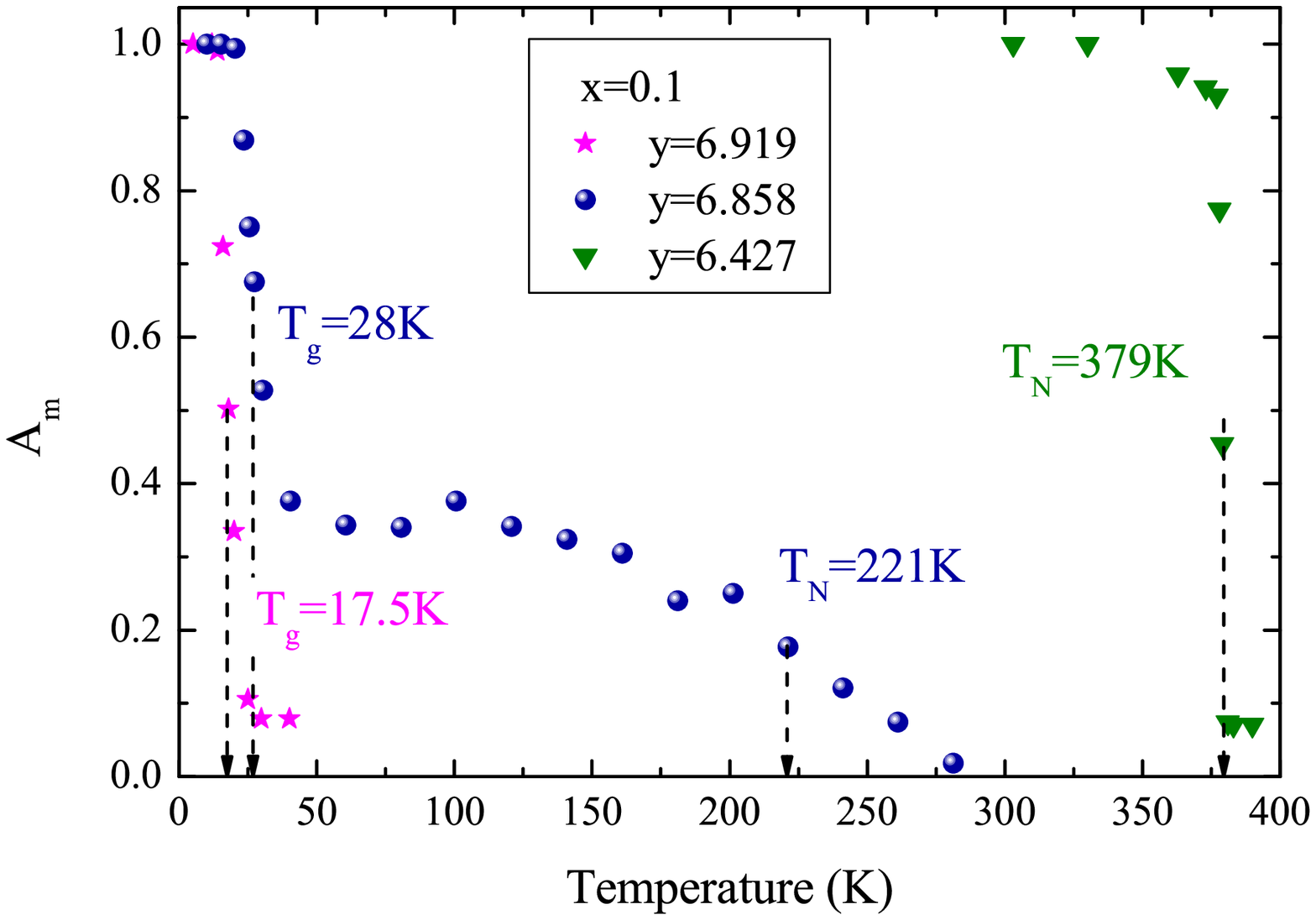}%
\caption{The magnetic amplitude $A_{m}$
fraction extracted from the muon depolarization as a function of temperature
for the 3 samples shown in Fig.~\protect\ref{musrzfraw}.}%
\label{am(t)}%
\end{center}
\end{figure}

In order to determine the magnetic critical transition temperatures, the
data were fitted to a sum of two functions: a Gaussian with amplitude $A_{n}$
representing the normal fraction of the sample, and a rapidly relaxing and
oscillating function, with amplitude $A_{m}$ and angular frequency $\omega $%
, representing the magnetic fraction and describes the magnetic field due to
frozen electronic moments \cite{OferPRB06,OferPRB08}. In this function the
sum $A_{m}+A_{n}=1$ is constant at all temperatures. Figure \ref{am(t)}
shows $A_{m}$ as a function of temperature, for the three samples in Fig.~%
\ref{musrzfraw}. Above the transition, where only nuclear moments
contribute, $A_{m}$ is close to zero. As the temperature decreases, the
frozen magnetic part increases and so does $A_{m}$, at the expense of $A_{n}$%
. For the pure AFM and SG phases, the transition temperature was determined
as the temperature at which $A_{m}$ is half of the saturation value. For the
samples with two transitions, two temperatures were determined using the
same principle.

The magnetic order parameter is extracted from the muon rotation frequency
in zero field as seen in Fig.~\ref{musrzfraw}(a). These rotations allow us
to determine $\omega(T,x,y)$. The temperature dependence of $\omega$ is used
to determine the effective interlayer and anisotropic interaction $%
\alpha_{eff}$ in Fig.~\ref{alpha}. $\omega(0,x,y)$ is used to determine the
family-dependent critical doping in Fig. \ref{unified}(b).

\section{Pseudogap\label{PG}}

\begin{figure}
[ptb]
\begin{center}
\includegraphics[
natheight=8.655900in,
natwidth=6.215400in,
height=2.9888in,
width=2.1534in
]%
{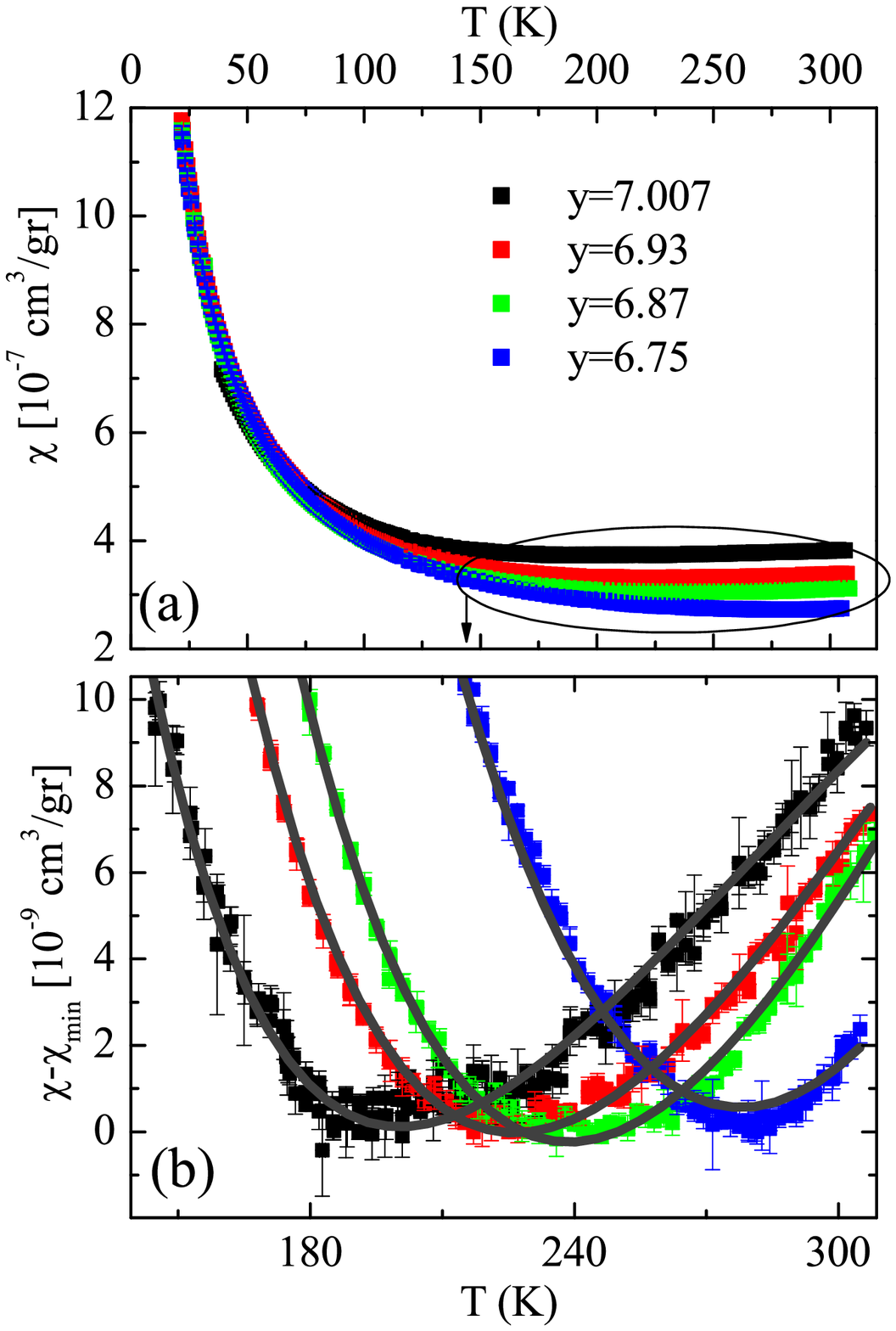}%
\caption{(a) Raw susceptibility data
from four samples of the $x=0.2$ family with different doping levels. (b)
Zoom-in on the data in the ellipse of panel a after the minimum value of $%
\protect\chi$ is subtracted. The solid lines are fits described in the text.}%
\label{chi}%
\end{center}
\end{figure}

We determine $T^{\ast}$ using temperature-dependent magnetization
measurements. In Fig.~\ref{chi}(a) we present raw data from four samples of
the $x=0.2$ family with different doping levels. At first glance the data
contain only two features: A Curie-Weiss (CW) type increase of $\chi$ at low
temperatures, and a non-zero baseline at high temperature ($\sim300$). This
base line increases with increasing y. The CW term is very interesting but
will not be discussed further here. The baseline shift could be a
consequence of variations in the core and Van Vleck electron contribution or
an increasing density of states at the Fermi level.

A zoom-in on the high temperature region, marked by the ellipse, reveals a
third feature in the data: a minimum point of $\chi$. To present this
minimum clearly we subtracted from the raw data the minimal value of the
susceptibility $\chi_{min}$ for each sample, and plotted the result on a
tighter scale in Fig.~\ref{chi}(b). The $\chi$ minimum is a result of
decreasing susceptibility upon cooling from room temperature, followed by an
increase in the susceptibility due to the CW term at low $T$. This
phenomenon was previously noticed by Johnston in YBCO \cite{JohnstonBook},
and Johnston \cite{JohnstonPRL89} and Nakano \cite{NakanoPRB94} \emph{et al.}
in La$_{2-x}$Sr$_{x}$CuO$_{4}$ (LSCO). The minimum point moves to higher
temperatures with decreasing oxygen level as expected from $T^{\ast}$. There
are three possible reasons for this decreasing susceptibility: (I)
increasing AFM correlations upon cooling, (II) opening of a SG where
excitations move from $q=0$ to the AFM wave vector \cite{ChiExp}, or (III)
disappearing density of states at the Fermi level as parts of the Fermi arc
are being gapped out when the PG opens as $T/T^{\ast}$ decreases \cite%
{KanigelNature06}.

In order to determine the $T^{\ast}$ we fit the data to a three-component
function $\chi=C_{1}/(T+\theta)+C_{2}\tanh(T^{\star}/T)+C_{3}$. The fits are
presented by solid lines in Fig.~\ref{chi}. The values of $T^{\star}$ are
shown in Fig.~\ref{criticalvsy} and Fig.~\ref{unified}.

\section{Doping and Impurities\label{Impurities}}

\begin{figure}
[ptb]
\begin{center}
\includegraphics[
natheight=8.655900in,
natwidth=6.215400in,
height=2.9888in,
width=2.1534in
]%
{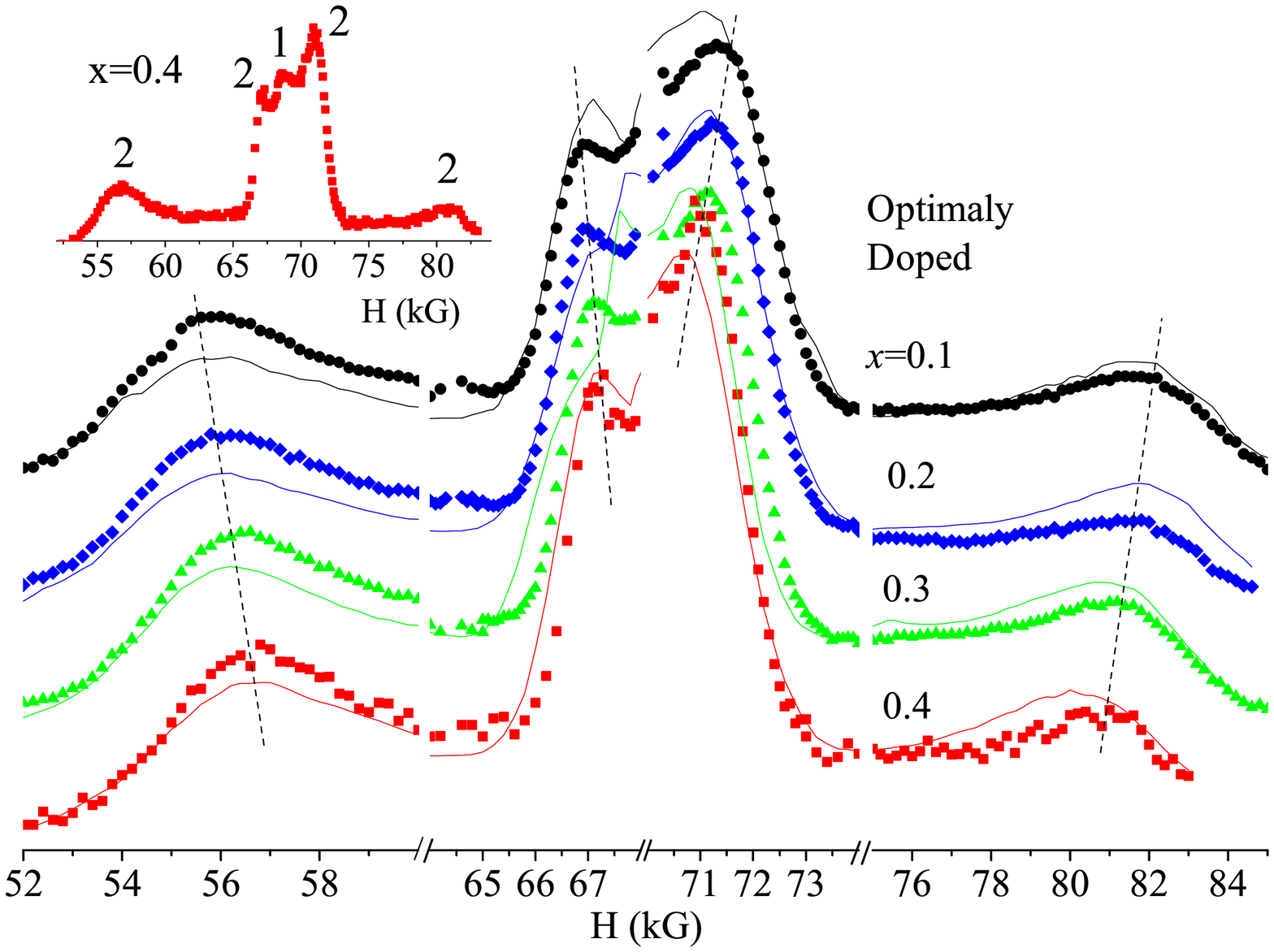}%
\caption{NMR spectra of $^{63}$Cu at $%
T=100$ K in optimally doped CLBLCO samples with varying $x$. The inset shows
the full spectrum of the $x=0.4$ compound including contributions from the
chain copper Cu(1) and plane copper Cu(2). The main figure zooms in on the
Cu(2) contribution (note the three axis breakers). The position of the Cu(2)
peaks are shown by dotted lines. The solid lines are fits to NQR line shape,
out of which $\protect\nu _{Q}$ and $\Delta \protect\nu _{Q}$ of Fig.~%
\protect\ref{nqranalysis} are obtained \protect\cite{KerenPRB06}.}%
\label{nqrline}%
\end{center}
\end{figure}

The Cu-NQR experiment is done on powder samples fully enriched with $^{63}$%
Cu. We measured between five and seven different samples for each $x$ in the
normal state at 100 K. The most overdoped sample is a non superconducting $%
x=0.1$ compound. The NMR measurements were done by sweeping the field in a
constant applied frequency $f_{app}$=77.95 MHz, using a $\pi/2$-$\pi$ echo
sequence. The echo signal was averaged 100,000 times and its area evaluated
as a function of field. The data are presented in Fig.~\ref{nqrline}. The
full spectrum of the optimally doped $x=0.4$ sample ($y=7.156$) is shown in
the inset of Fig.~\ref{nqrline}. The main planes emphasize the important
parts of the spectrum using three axis breakers.

The evolution of the main peaks as $x$ increases is highlighted by the
dotted lines. It is clear that as $x$ decreases the peaks move away from
each other. This means that $\nu _{Q}$ at optimal doping is a decreasing
function of $x$. A more interesting observation is the fact that there is no
change in the width of the peaks, at least not one that can easily be
spotted by the naked eye. This means that the distribution of $\nu _{Q}$ is $%
x$-independent and that there is no difference in the disorder between the
optimally doped samples of the different families. Thus, as mentioned in
Sec.~\ref{Main}, disorder is not relevant to the variation of $T_{c}^{max}$
between the different families. This conclusion is supported by more
rigorous analysis \cite{KerenPRB06}. $\nu _{Q}$ and $\Delta \nu _{Q}$
presented in Fig.~\ref{nqranalysis} are obtained by fitting NQR line shape
to this data as demonstrated by the solid lines.

\section{Penetration depth\label{PenDep}}

\begin{figure}
[ptb]
\begin{center}
\includegraphics[
natheight=8.085100in,
natwidth=6.203300in,
height=2.9888in,
width=2.2995in
]%
{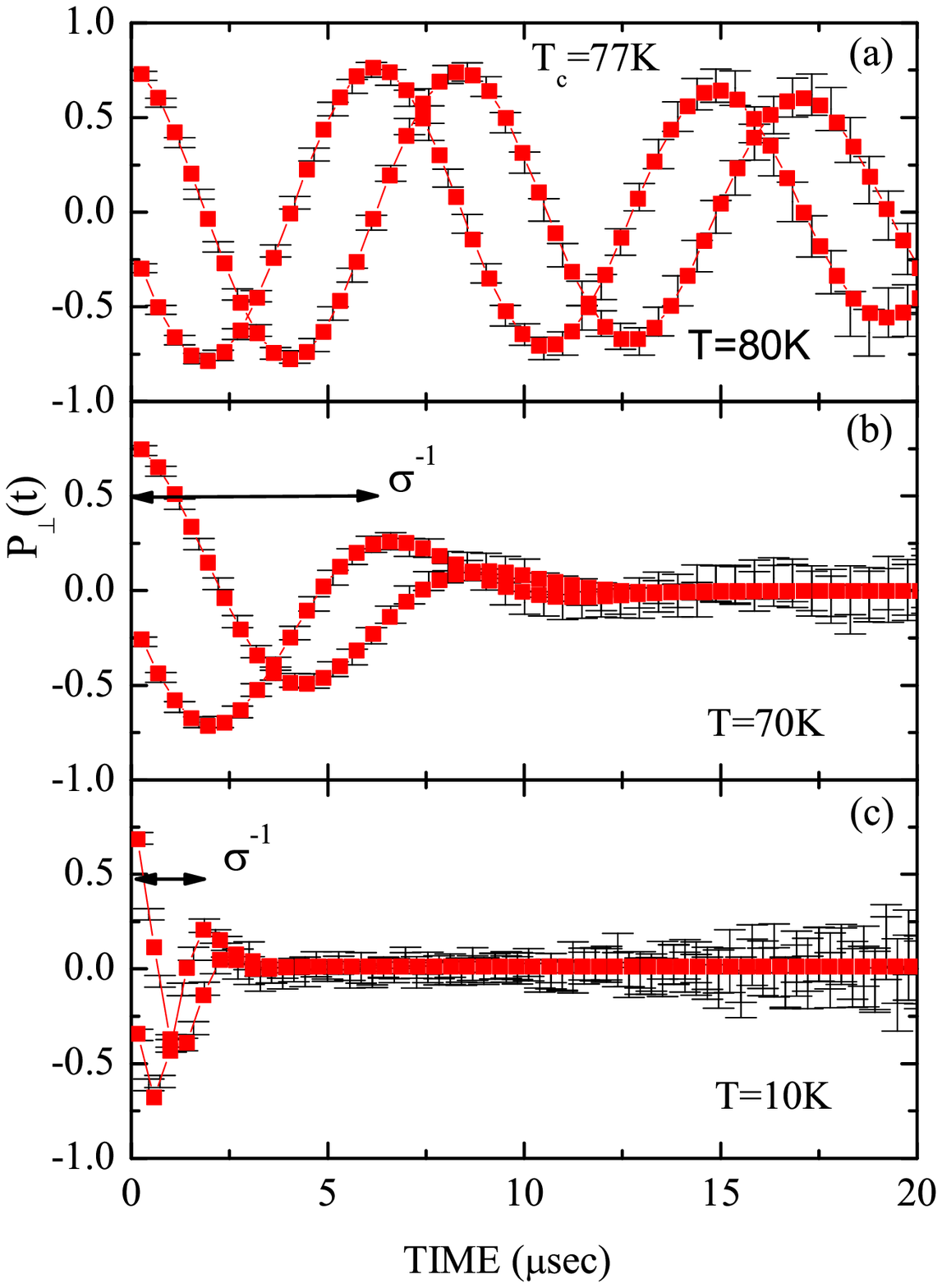}%
\caption{Demonstrating the real and
imaginary polarization in a TF-$\mu$SR experiment as the temperautre is
lowered below $T_{c}$. The data are presented in a rotating reference frame.}%
\label{musrtfraw}%
\end{center}
\end{figure}

The penetration depth is measured with transverse field TF-$\mu$SR. In this
experiment one follows the transverse muon polarization $P_{\bot}(0)$ when a
magnetic field is on and perpendicular to the initial polarization. These
experiments are done by field cooling (FC) the sample to 1.8 K at an
external field of 3 kOe. Every muon precesses according to the local field
in its environment. When field cooling the sample, a vortex lattice is
formed, and the field from these vortices decays on a length scale of the
penetration depth $\lambda$. This leads to an inhomogeneous field
distribution in the sample. Since the magnetic length scale is much larger
than the atomic one, the muons probe the magnetic field distribution
randomly, which, in turn, leads to a damping of the muons average spin
polarization. This situation is demonstrated in Fig.~\ref{musrtfraw} where
we present $P_{\bot}(0)$ in two different perpendicular directions (called
real and imaginary) in a rotating reference frame. At temperatures above $%
T_{c}$ the field is homogeneous and all muons experience the same field, and
therefore no relaxation is observed. Well below $T_{c}$ (of $77$~K in this
case) there are strong field variations and therefore different muons
precess with different frequencies, and the average polarization quickly
decays to zero. At intermediate temperatures the field variation are not
severe and the relaxation is moderate.

It was shown that in powder samples of HTSC the muon polarization $P_{\bot
}(t)$ is well described by $P_{\bot}(t)=exp(-R_{\mu}^{2}t^{2}/2)cos(\omega
t) $ where $\omega=\gamma\mu H$ is the precession frequency of the muon, and 
$R_{\mu}$ is the relaxation rate \cite{MuonBook}. The solid line in this
figure is the fit result. The fact that the whole asymmetry relaxes
indicates that CLBLCO is a bulk superconductor. The fit results for $R_{\mu}$
are shown in Fig. \ref{Uemura}. As can be seen, the dependence of $T_{c}$ on 
$R_{\mu}$ is linear in the under-doped region and universal for all CLBLCO
families, as expected from the Uemura relations. However, there is a new
aspect in this plot. There is no "boomerang" effect, namely, overdoped and
underdoped samples with equal $T_{c}$ have the same $R_{\mu}$, with only
slight deviations for the $x=0.1$. Therefore, in CLBLCO there is one to one
correspondence between $T_{c}$ and $R_{\mu}$, and therefore $n_{s}/m^{\ast}$%
, over the whole doping range.

\section{Lattice parameters\label{Lattice}}

\begin{figure}
[ptb]
\begin{center}
\includegraphics[
natheight=8.943000in,
natwidth=6.602000in,
height=2.9897in,
width=2.2148in
]%
{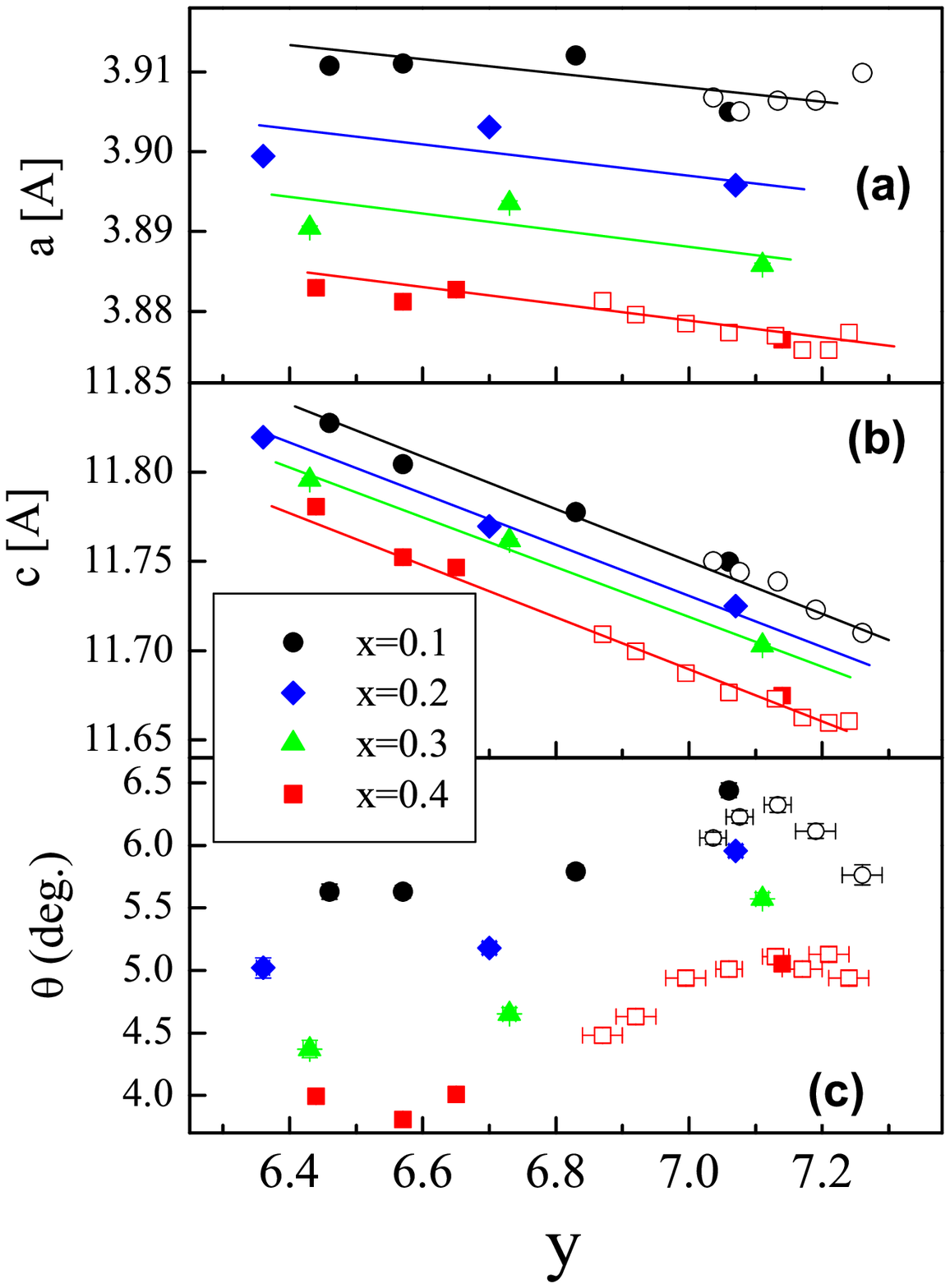}%
\caption{Lattice parameters as a
function of oxygen doping. (a) The lattice parameter $a$. (b) The lattice
parameter $c$. (c) $\protect\theta$ - the buckling angle between the copper
and oxygen in the plane. The empty symbols are from Ref.~\protect\cite%
{ChmaissemNature99}. The lines are guides to the eye.}%
\label{neutrons}%
\end{center}
\end{figure}

Neutron powder diffraction experiments were performed at the Special
Environment Powder Diffractometer at Argonne's Intense Pulsed Neutron Source
(see Ref.~\cite{ChmaissemNature99} for more details). Figure \ref{neutrons}
shows a summary of the lattice parameters. The empty symbols represent data
taken from Ref. \cite{ChmaissemNature99}. All the parameters are
family-dependent, but not to the same extent. The lattice parameters a and
c, depicted in Fig. \ref{neutrons}(a) and (b), change by up to about 0.5\%
between the two extreme families ($x=0.1$ and $x=0.4$). The in-plane Cu-O-Cu
buckling angle is shown in Fig. \ref{neutrons}(c). This angle is non-zero
since the oxygen is slightly out of the Cu plane and closer to the Y site of
the YBCO structure. As mentioned in Sec.~\ref{Main} it changes by 30\%
between families.

\end{document}